# Optimal Operation of Interdependent Power Systems and Electrified Transportation Networks

M. Hadi Amini[1,2,3] and Orkun Karabasoglu[1,3,4*]

[1]*SYSU-CMU Joint Institute of Engineering, Guangzhou, China*
[2]*Department of Electrical and Computer Engineering, Carnegie Mellon University, Pittsburgh, PA, US*
[3]*School of Electronics and Information Technology, SYSU, Guangzhou, China*
[4]*Scott Institute for Energy Innovation, Carnegie Mellon University, Pittsburgh, PA, US*
<u>amini@cmu.edu</u>, <u>karabasoglu@cmu.edu</u>

***Abstract:*** Electrified transportation and power systems are mutually coupled networks. In this paper, a novel framework is developed for interdependent power and transportation networks. Our approach constitutes solving an iterative least cost vehicle routing process, which utilizes the communication of electrified vehicles (EVs) with competing charging stations to exchange data, such as electricity price, energy demand, and time of arrival. EV routing problem is solved to minimize the total cost of travel using the Dijkstra algorithm with the input from EVs battery management system, electricity price from charging stations, powertrain component efficiencies, and transportation network traffic conditions. Through the bidirectional communication of EVs with competing charging stations, EVs charging demand estimation is done much more accurately. Then, the optimal power flow problem is solved for the power system, to find the locational marginal price at load buses where charging stations are connected. Finally, the electricity prices are communicated from the charging stations to the EVs, and the loop is closed. Locational electricity price acts as the shared parameter between the two optimization problems, i.e. optimal power flow and optimal routing problem. Electricity price depends on the power demand, which is affected by the charging of EVs. On the other hand, location of EV charging stations and their different pricing strategies might affect the routing decisions of the EVs. Our novel approach that combines the electrified transportation with power system operation, holds tremendous potential for solving electrified transportation issues and reducing energy costs. The effectiveness of the proposed approach is demonstrated using Shanghai transportation network and IEEE 9-bus test system. The results verify the cost-savings for both power system and transportation networks.

---

# 1. Introduction

## 1.1. Motivation

Electrified vehicles (EVs) can help reducing national gasoline consumption, greenhouse gas emissions and dependency on foreign oil since they operate partly or entirely on inexpensive electricity. Electricity needed to charge the EVs can be potentially obtained from local, renewable, and less carbon-intensive energy sources [1],[2]. However, incorporating large number of EVs into existing power and transportation infrastructures remains to be a challenging task [3]. According to [4], China has increased the EV utilization vigorously since 2009. The target production and sales numbers of Battery-only EVs and Plug-in Hybrid EVs are expected to be 5 million until 2020 [4]. This ever-increasing use of electric vehicles and charging stations, may affect both power and transportation networks.

Electrified transportation networks and power systems are mutually coupled networks. In the interdependent power and transportation systems literature, transportation network dynamics have not been comprehensively considered. For instance, electricity price is a function of charging demand, which is a transportation based parameter. Similarly, optimal power system operation might have a significant impact on transportation, i.e. the availability of battery energy is a function of charging transactions. Battery energy in EVs determines the electric range impacting the transportation network dynamics. At the same time, location of EV parking lots that serve as charging stations also has the potential to change the routes of EVs that need to be charged. Therefore, power and transportation networks should be controlled and optimized together to operate optimally because the decoupled optimization of each standalone network might lead to locally optimum operation. In other words, joint optimization might lead to a globally optimum operation and cost reductions. In this study, we aim to develop a comprehensive framework and methodologies that consider the constraints of both networks to achieve the joint optimal operation. Following, we review the related literature and present our contributions.

## 1.2. Literature Review

Relevant work can be classified into: (1) EV charging demand estimation, (2) optimal power systems operation with large-scale penetration of EVs, (3) electrified transportation and route optimization, and (4) interdependent networks. We review each category with detail in the following paragraphs. Finally, we explain our contributions.

Increased energy demand due to EV charging might lead to unwanted peaks in the electricity consumption [5]. Clement-Nyns et al. studied the effect of EV charging on a residential distribution grid in [5], but the effect of EV charging on transportation networks is not addressed. Zhang and Grijalva [6] proposed a data driven queuing model for residential EV charging demand by performing big data analytics on smart meter measurements. Alizadeh et al. [7] proposed a stochastic model based on queuing theory for the electric vehicle charging demand. Deilami et al. [8] worked on a real-time load management solution for coordinating the charging of multiple plug-in electric vehicles in a smart grid system. Mingfei and Jilai [9] studied the regular behavior process of the private electric vehicle clusters to model the EVs charging demand. Donadee et al. [10] developed a method for optimal autonomous charging of EVs for the estimated energy demand that is based on collected driving patterns. Power system operation requires accurate estimation of the increased energy demand due to EV charging. Hence, quantification of the EV charging demand is required to help power system operators obtain more optimal solutions.

EVs have been considered as advantageous resources to improve power system operation [11]. They can act as adjustable energy storage resources and participate in ancillary services procurement [12]. Hence, EV can provide distributed energy storages for power systems [13]. In this context, Yazdani et al. studied the impact of EVs as distributed energy storages on the electricity demand of smart grids [14]. In [15], Amini et al. developed an accurate EV charging demand forecaster for stochastic optimal operation of power systems. In [16], a reliability constrained optimization problem is modeled for allocation of EV parking lots in distribution networks. A simultaneous optimal allocation framework for EV charging stations and renewable resources is presented in [3]. In [17] a two-stage model to allocate EV parking lots in distribution systems considering power loss, network reliability, and voltage deviation is presented. Darabi et al. [18] provided a thorough analysis on the plug-in hybrid electric vehicles historical driving data from the National Household Travel Surveys (NHTS). In [19] demand response strategies are incorporated to offer financial benefits for EVs to optimize their charging fees. Although EVs can ameliorate the performance of power systems, they are mobile loads and this should be taken into account while studying the large scale integration of EVs.

Electric vehicles can be seen as mobile loads on a geographic region that will eventually connect to power system via charging stations. Vehicle routing is the process that connects vehicles with their destinations but also with the power system. There are works that focus on optimal routing through shortest path in conventional vehicles [20]-[23]. Traditionally, route optimization for conventional vehicles has been done through solving shortest path problems. For the solution of shortest path problems, different algorithms have been proposed such as Dijkstra's Algorithm [20], Genetic Algorithms [21], Improved Bellman–Ford Successive Approximation Algorithm [22], Particle Swarm Optimization [23] and Column Generation techniques [24]. Shortest path does not guarantee the least cost path [25]. There are other factors including traffic, driving patterns, terrain, vehicle load, and air conditioner load, which affect the efficiency of the vehicle and change the cost of travel. Karabasoglu and Michalek [26] investigated the effect of driving patterns on the life cycle cost and emissions of different categories of vehicle powertrains under various scenarios and simulated driving conditions. They found that driving patterns matter and have the potential to change the ranking of advanced powertrains for their benefits. Zhang [27] integrated information of road terrain with the energy management systems of hybrid vehicles and showed that the incorporation of terrain review can help reducing the fuel consumption in electric vehicles. Xiao et al. [28] investigated the impact of vehicle load on the classical capacitated vehicle routing problem and found that load is an important factor to consider. Tavares et al. [29] showed that vehicle weight and the inclination of roads affects the efficiency of the vehicle. Their routing strategy increased cost savings by 8%. Artmeier et al. [30] proposed extensions to general shortest-path algorithms, which addressed the problem of energy-based-optimal routing in the presence of rechargeable batteries. Qiao and Karabasoglu [25] proposed a novel framework for vehicle powertrain connected route optimization which considers dynamically changing conversion factors with various traffic situations and other sensory information from powertrain components such as motor, engine and battery during route optimization. Li et al. [50] tackled the problem of risk-averse route planning in a transportation network with time-dependent and stochastic costs.

There are studies that focused on power system optimization problem while considering the impact of EVs on the electricity demand. Yang et al. in [31] proposed a distributed mixed optimization approach to solve the joint scheduling problem of large-scale smart appliances and batteries. Their objective is to minimize electricity payment, user's dissatisfaction and battery loss. Furthermore, they formulated the battery scheduling problem as a mixed-integer linear program which is solved using Benders decomposition technique. In [32], You et al. moved one step further and proposed a model of a battery switching station (BSS) for electric buses (EBs) to solve the EV battery charging scheduling problem. In their model, each EB determines a battery available for switching. In order to compute the optimal schedule, they utilized the dual decomposition to decouple the charging decisions at various charging boxes [32]. The main advantage of our proposed method compared with this study is considering the transportation network model by deploying the optimal routing to determine the EV charging demand. Bashash et al. evaluated the charge pattern optimization of plug-in EVs by defining the timing and rate of obtaining electricity from the power systems. Although they minimized the total cost of electricity and gas as well as the battery health degradation simultaneously [33], the interdependent effects of power and transportation networks are not captured in their proposed solution. Xiong et al. proposed a systematic simulation framework to analyze the effect of Plug-in EV charging stations on the power distribution system and transportation networks [34].

There are studies that focused on interdependent power and transportation networks. A comprehensive review of available benchmarks for TE simulation is provided in [35]. Viswanath and Farid proposed a holistic transportation-electricity nexus to model the coupled kinematic and electrical states. The proposed hybrid dynamic system model was built on marked petri-net model with the continuous time kinematic and electrical state evolution [36]. Farid in [37] developed a hybrid dynamic system model for transportation electrification. His proposed model included multimodality and multiagency-based transportation network models. Similar to [36], a Petri-net model superimposed on the continuous time kinematic and electrical state evolution was deployed for the hybrid dynamic system modeling purpose. Farid in [38], argued the requirement for a large scale TE test case, and proposed Symmetrica for modern transportation electrification studies. To this end, Farid defined transportation-electricity Nexus as a system-of-systems including two benchmarks for power system topology and transportation network.

### 1.3. Contribution of this paper

In this paper, we introduce a novel framework and methodology to integrate electrified transportation with power system for globally optimum operation that reduces costs. Compared to the previous aforementioned studies, we propose a bidirectional information exchange mechanism between power and transportation networks by modeling the interaction of EVs and charging stations; develop a novel routing strategy that considers location and electricity price of charging stations, and changing efficiencies of EVs under different traffic situations; and optimize the power system operation with the more accurate electricity demand of EVs and finally feeding the electricity prices back to charging stations in the transportation network which closes the loop. We also implement our methodology using a more realistic model of transportation network with different traffic information and a power network topology. Fig. 1 represents the general overview of the proposed framework. We explain the proposed framework in Section 2.

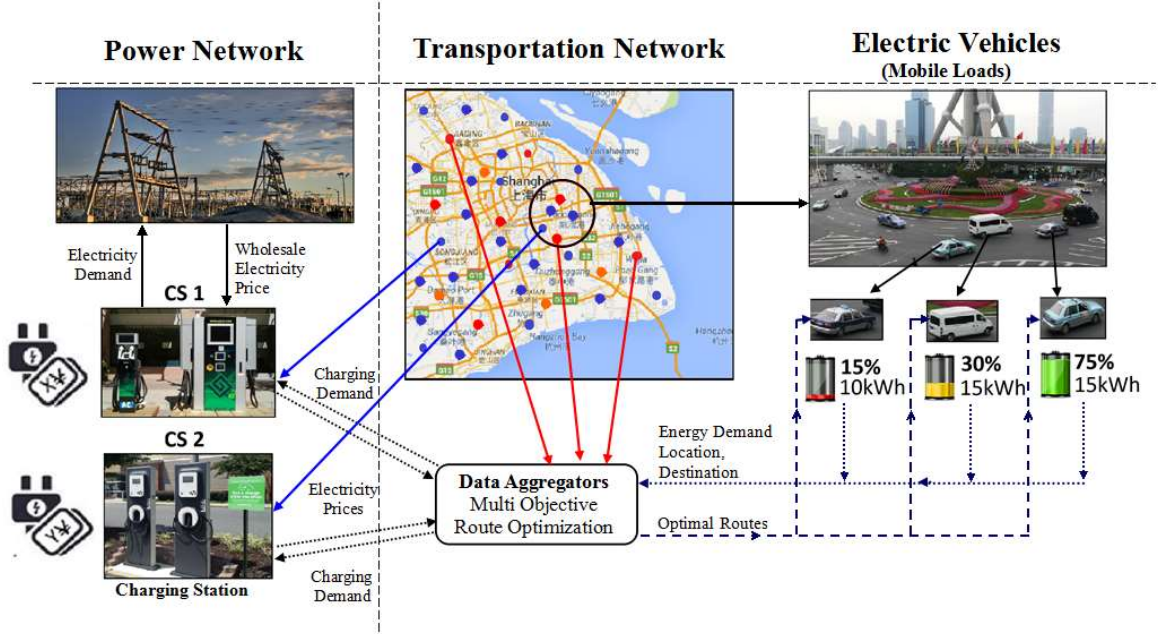

**Fig. 1. General framework for simultaneous operation of electrified transportation and power networks**

The main contribution of this paper is threefold:

(1) We propose a novel framework for coupled electrified transportation and power networks to achieve optimal operation. We consider the shared variables between the interdependent networks to evaluate the interaction of two networks;

(2) We develop a least cost routing strategy for electric vehicles to connect them with competing charging stations and their destinations: Our routing strategy considers charging costs, transportation cost, and traffic situation;

(3) We analyze the cost benefits of the proposed framework by realistic simulations using a model of Shanghai transportation and power network.

### 1.4. Organization of the Paper

The rest of this paper is organized as follows. Section 2 presents the methodology for the interdependent operation of power and transportation networks. We introduce least cost routing optimization and the details of optimal power flow to determine the locational marginal prices. Case study and simulation results are provided in Section 3. In Section 4 summary and conclusions are given.

## 2. Methodology

In this section, we elaborately provide a description of the interdependent nature of power and transportation networks. As we have shown in Fig. 1, the transportation network includes the electric vehicles, charging stations and the proposed data aggregators as its basic components. The data aggregators collect the vital information such as the current location, destination, battery size, state of charge, and routing preferences from the electric vehicles as well as the location, and price of electricity from the charging stations. The data aggregators send the charging load demand to the charging stations and the optimal routing solutions to the electric vehicles once the routing problem is solved on the cloud. The charging stations send the total electricity demand to the power network and get the wholesale electricity prices in return. The optimal routes are provided to the electric vehicles based on the traffic and the electricity price at any particular charging station. Our framework not only looks into minimizing the transportation cost for the electric vehicles but also considers the charging cost. Least cost routing serves as the function that connects EVs (mobile loads on transportation network) with charging stations on power network. By connecting the interdependent networks, the system can estimate the electricity demand for each charging station more accurately and in advance before it is realized so that the power system can optimally balance power generation and consumption to provide a more reliable service.

Here, we summarize three key problems that are addressed in our paper:

- EVs have stochastic nature on the transportation network and it is difficult to accurately estimate EV charging load for power network while they are mobile.
- Electric vehicle routing decisions are difficult considering limited energy storage, traffic conditions and multiple charging stations with different price alternatives.
- Linking the electrified transportation network with the power network is complex and it is difficult to optimize the operation of both networks.

To solve the mentioned issues, we formulate a two-stage optimization problem to perform the route optimization and the power system optimization simultaneously.

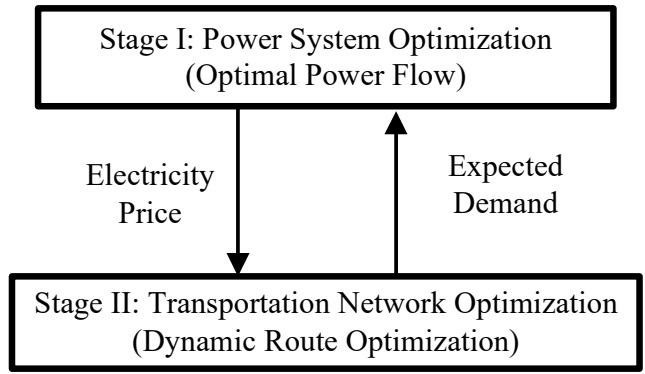

**Fig. 2. General two-stage optimization problem**

## 2.1. Integration of Interdependent Power and Transportation Networks

The general framework of two-stage optimization problem is shown in Fig.2. As illustrated in the figure, electricity prices are communicated from electric power network to the transportation network. Furthermore, the expected charging demand of each EV charging station is communicated to the power network for updating the results of optimal power flow problem. A more detailed model of our proposed framework is represented in Fig. 3. According to this figure, physical network layer includes three main elements and one sub-level element. Power generation units, conventional load demand (before adding parking lots with EV charging capability), and charging stations connected to power network are the main elements. EVs are considered as the sub-level element. After the charging stations receive the locational marginal electricity price from the independent system operator (ISO), send the charging price based on their incentivizing strategy to EVs. The transportation layer is coupled with power network via EVs with different sizes of battery packs. The price signals and EV demand at each charging station are the shared variables. Once, EVs receive the electricity prices from the charging stations, cost-optimal routing is done using traffic information, battery initial state of charge, battery size, and changing EV powertrain efficiency under different traffic conditions.

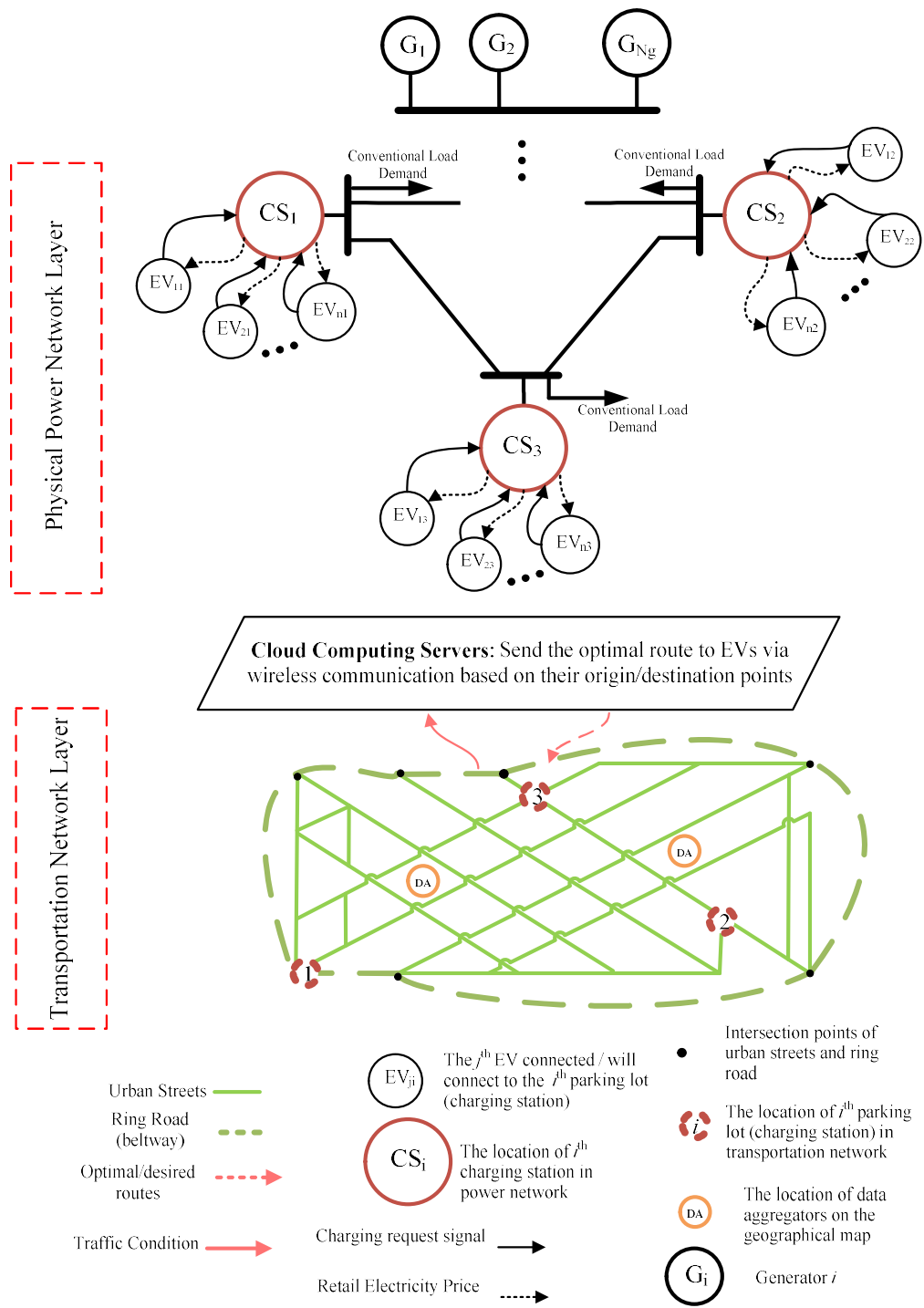

**Fig. 3.** Schematic view of the proposed framework for the optimal operation of transportation and power networks

## 2.2. An Information Exchange Mechanism for EVs and Charging Stations

Algorithms 1 and 2 represent the detailed procedure for the charging stations and EVs to determine the charging electricity price and the optimal route. Note that the EVs have a two stage routing optimization: 1) Primary one-level routing from starting point to the destination, 2) Secondary bi-level routing optimization to determine the optimal routes from current location to the charging station and from the charging station to the destination.

---

*Algorithm 1: Interaction of the charging stations with EVs to determine the price signals*

---

**1.** EV charging stations receive the locational marginal prices from the ISO

**2.** EV charging stations receive the signal from each potential EV via cloud computing server whether the EV wants to potentially charge its battery at the charging station

**3.** EV charging station receives expected duration, amount of charge required, and distance from each EV

**4.** EV charging station determines the electricity price for each EV and communicates the price

- go to {*Algorithm 2*}

**5.** Charging stations aggregate the expected EV charging demand based on the approved signals from EV to arrive at the charging station

**6.** Charging stations send their expected demands to the ISO for updating price signals for next time step

**7.** ISO solved DC optimal power flow (DCOPF) to determine the nodal electricity prices

**8.** Go to step 1

---

*Algorithm 2: Optimal routing of EVs utilizing bi-level routing scheme*

---

**1.** EVs receive the price signals from all charging stations via cloud computing server

**2.** Optimal routing system receives the traffic data from transportation network database

**3.** *Primary route optimization*: EVs receive the optimal route and send a charging request signal to the candidate charging stations

**4.** EVs receive the approval signal from the charging station

**5.** *Secondary route optimization*: EVs send the data set including their current location, destination, battery state of charge, and other required information to the cloud computing server for the purposes of bi-level optimal routing from the starting point to charging station and from the charging station to the destination

**6.** EVs receive the suggested optimal routes from cloud computing server

## 2.2.1. Electric Vehicle Module

The EV module represented in Fig. 4, includes the information flow inside each single EV and the data communication between charging stations and cloud EV data center. The *Routing Optimization Algorithm* which is mentioned in Fig. 4, will be elaborately introduced in Fig. 6. CPI stands for Charging station Profit-margin Index.

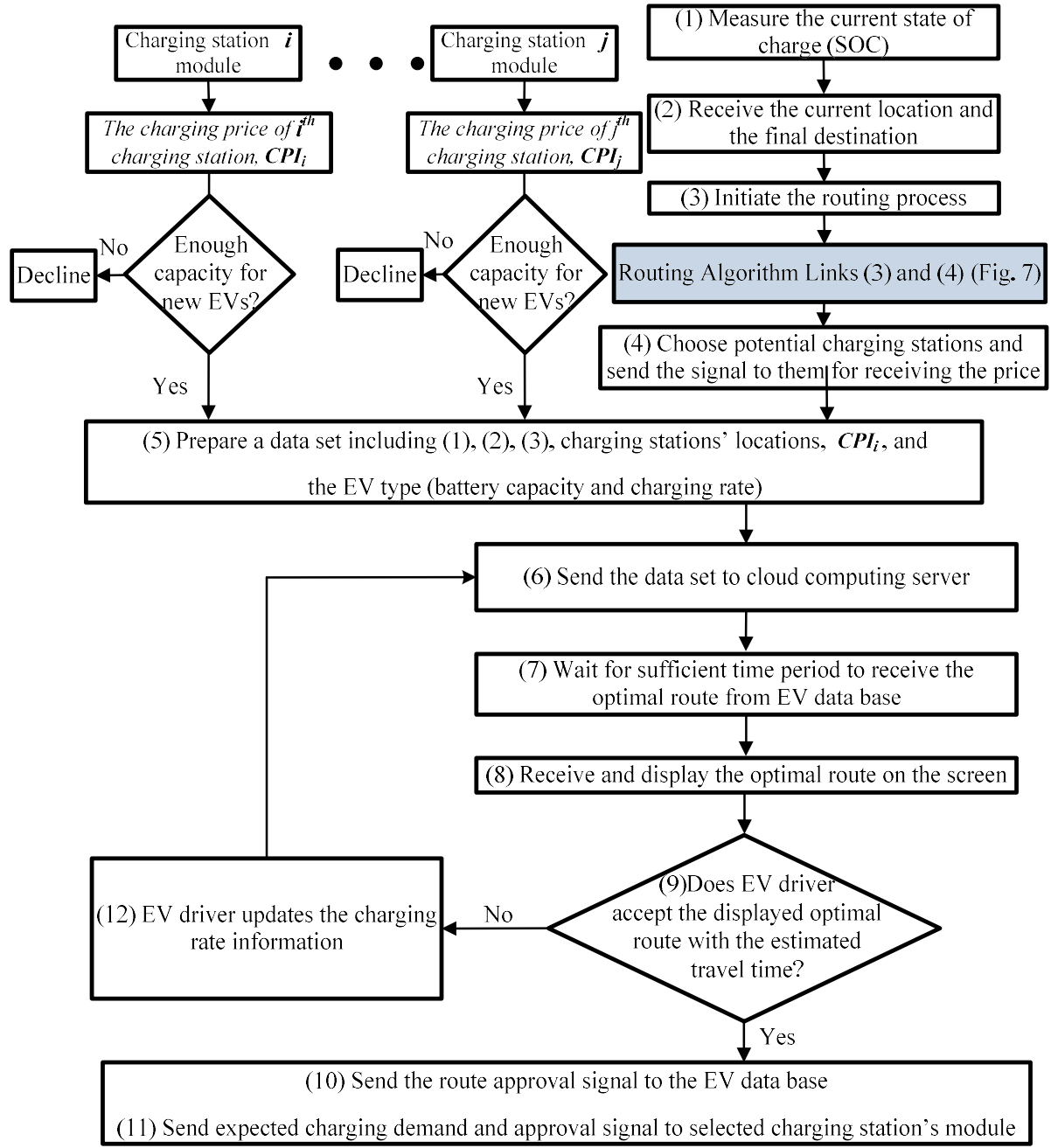

**Fig. 4. EV module representation**

### 2.2.2. Charging Station Module

This module receives the traffic information from the transportation network database and the electricity price signal from the power system operator, denoted by $p_m$. Fig. 5 represents the generalized charging station module. According to this module, EV charging stations receive the locational marginal prices from the ISO. Then, the EV charging stations receive the signal from each potential EV via cloud computing server whether the EV wants to potentially charge its battery at the charging station. In next step, EV charging station determines the electricity price for each EV and communicates the price. The price signals broadcast to EVs so that they can determine the optimal route using Charging Station Strategy-Vehicle Powertrain Connected Routing Optimization (CSS-VPCRO) method. Charging stations aggregate the expected EV charging demand based on the approved signals from EV to arrive at the charging station. They send their expected demands to the ISO for updating price signals for next time step (next hour).

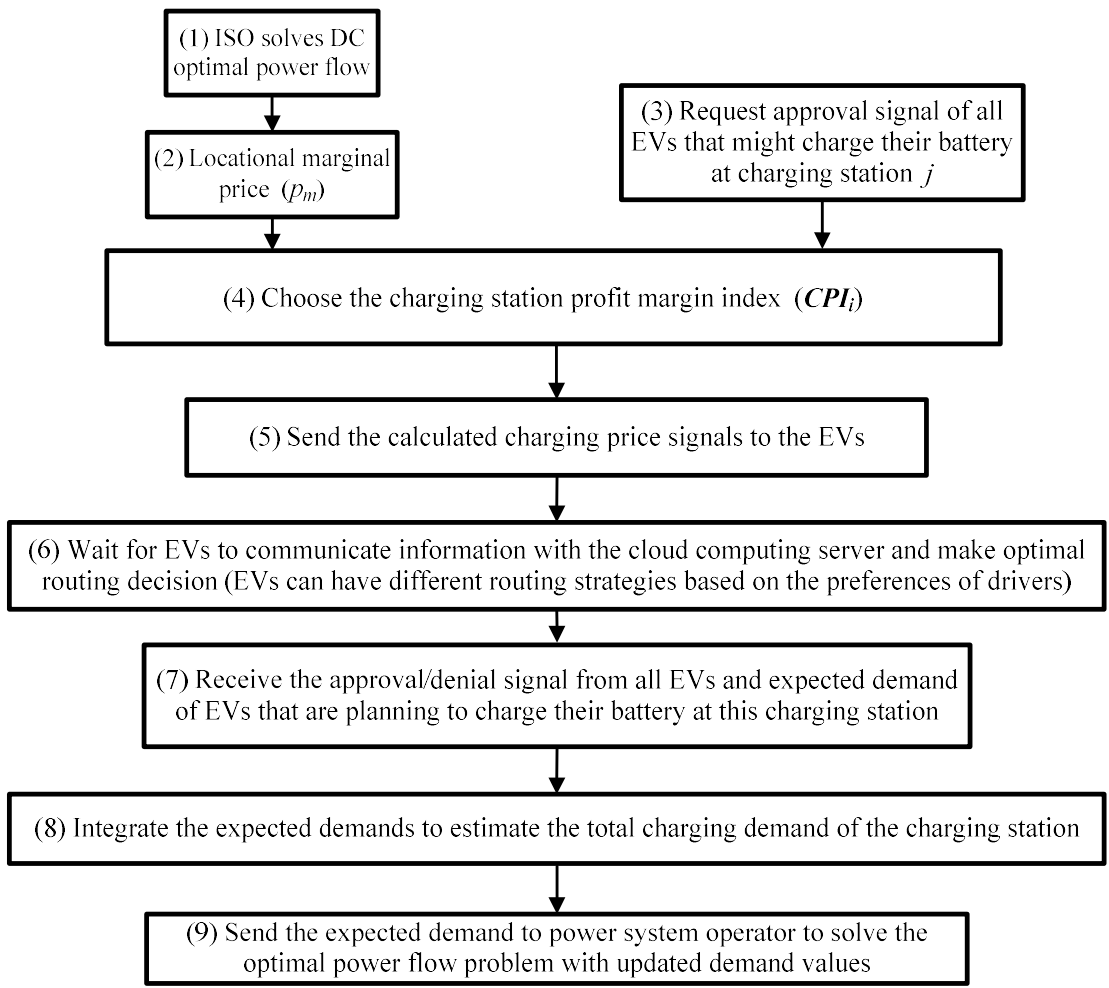

**Fig. 5. Charging station module representation**

We assume that each charging station uses a fixed Charging station Profit margin Index (*CPI*) which is determined by the charging station facility based on the number of charging requests, the location of charging station (downtown, vicinity, or residential area) and other factors of interest. We define the *Charging station Profit-margin Index (CPI)* to represent the profit margin of charging stations as the difference between the price determined by the charging station and the locational marginal price sent by ISO. This index shows the percentage of the price deviation for the profit of charging station. If the locational marginal price at the charging station bus is shown as $p_i$ and the offered price to EVs by the $i^{th}$ parking lot is denoted by $p_i^{CS}$, then the *Charging station Profit-margin Index*, denoted by $CPI_i$ can be calculated using the following equation:

$$CPI_i = \frac{p_i^{CS} - p_i}{p_i} \times 100 \tag{1}$$

Charging stations can apply different price incentives and change $p_i$ to maximize their profits depending on various factors such as capacity usage, parking space limitations, wholesale electricity price, and competition strategy against other charging stations, traffic scenarios, and arrival times.

**2.3. Vehicle Powertrain Connected Route Optimization Considering Charging Stations with Price Incentives**

In this section, we introduce the methodology for our proposed least cost optimal routing strategy for EVs considering Charging Stations with different price incentives. Our methodology builds on the vehicle powertrain connected routing optimization (VPCRO) proposed in [25]. Qiao and Karabasoglu proposed a novel routing approach considering the interaction between vehicle powertrains and traffic situations and their impact on the segment cost of the traffic network. They showed that shortest path is not necessarily the least-cost path and least cost paths can significantly change based on the powertrain type. If the vehicle powertrain is electrified, then battery capacity and initial battery state of charge also affect the least cost routes [25]. In this paper, we build on VPCRO and consider the bidding mechanism between EVs and charging stations while determining the least-cost route in a dynamically changing traffic network where network segments have different traffic levels (traffic levels change the efficiency of vehicle powertrain components, thus transportation cost). We refer to this approach as "Charging Station Strategy-Vehicle Powertrain Connected Routing Optimization (CSS-VPCRO)".

The percentage of the trips that will efficiently be covered on electricity before the need for a recharge or battery swap depends on the vehicle type, battery size, initial battery SOC, and traffic conditions [25]. The proposed routing strategy in this paper, illustrated in Fig. 6, takes into account all the aforementioned factors and provides least-cost optimal path. The Battery Management System (BMS) provides the battery information via the wired or wireless On Board Diagnostics (OBD) scanner that can access the information on the Control Area Network (CAN) bus of the vehicle. The data collection unit sends the vehicle data (location, destination, current state of charge, and battery size.) to the data aggregator. Data collection device can be a smart phone or GPS device. The data aggregator sends the vehicle and the transport network information to the cloud server and gets the electricity price and the optimized routes to the destination passing through the optimum charging station for each individual EV. This information is then sent to the EV's data collection unit and the driver is informed about the optimal routes.

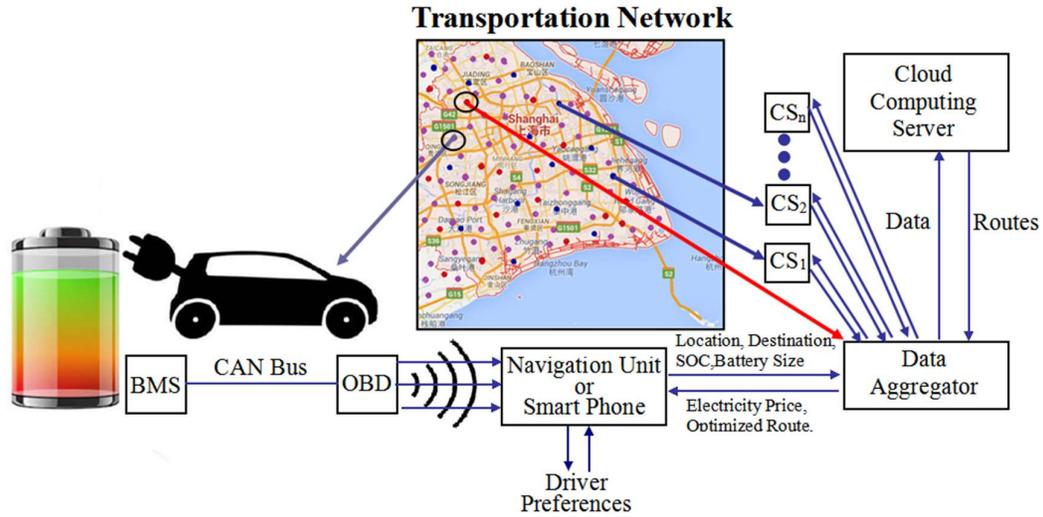

**Fig. 6. Least cost optimal routing strategy for EVs considering charging stations with different price incentives**

The routing process for electrified vehicles is illustrated in Fig. 7. After the origin and the destination are determined for each trip, the vehicle is examined to see if its battery carries enough energy for the intended trip. If there is enough energy, then the optimal path is found using VPCRO and noted as $P(i) = [n_1(i), n_2(i), ..., n_k(i)]$. Then for each node, $n_j(i)$ on the path, the nearest charging station $m_j$ is found. The algorithm chooses *n* charging stations with the least cost along the route. These options are presented to the driver and driver chooses one of them. If there is not enough charge in the battery to cover the whole trip, the vehicle should be charged either at the origin, or at one point during the trip before it runs out of the energy. Our algorithm determines the node at which the vehicle runs out of charge and offers the charging stations to the driver. Note that this *Routing Optimization Algorithm* is used in the EV module presented in Fig. 4.

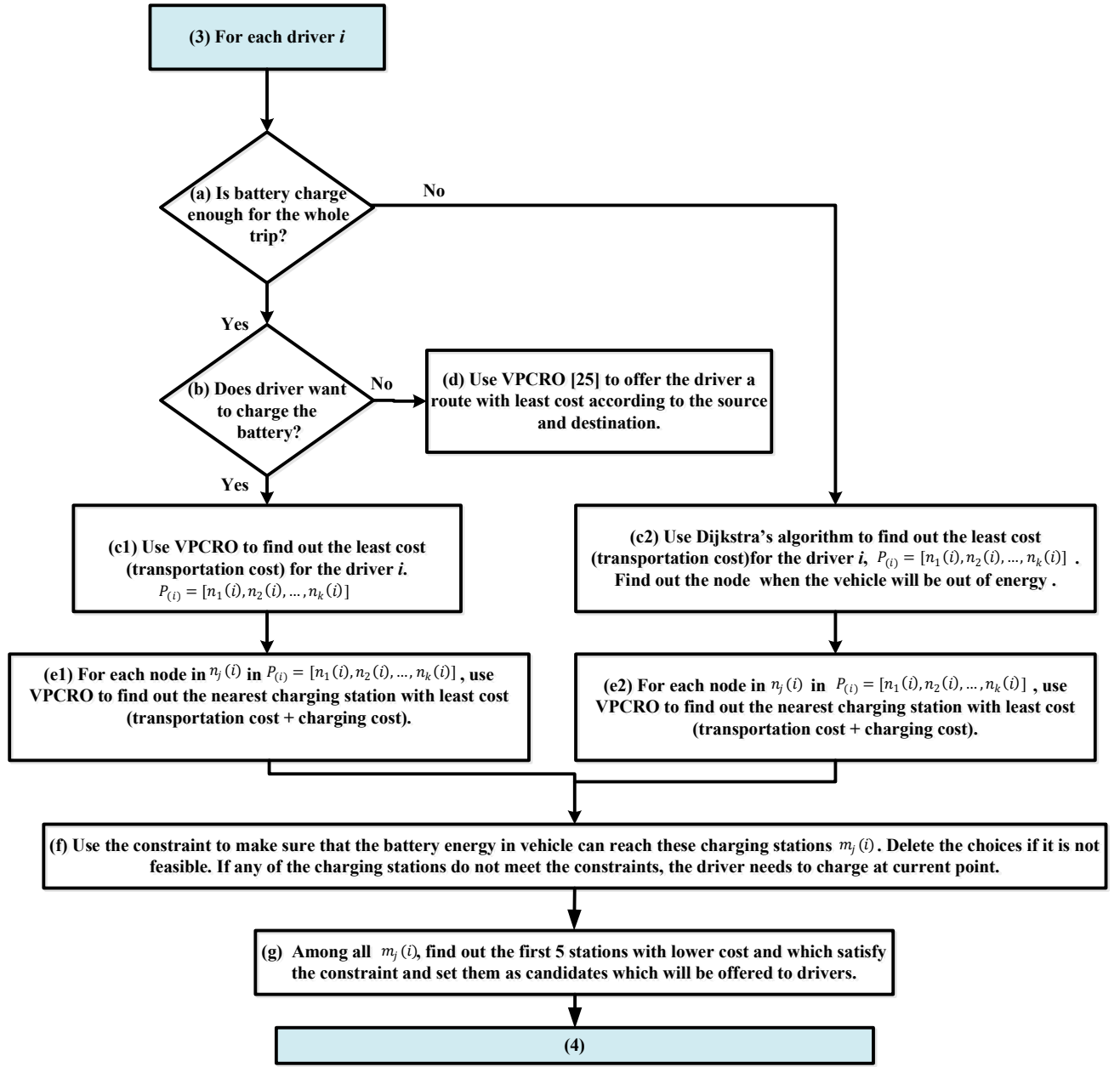

**Fig. 7. Routing optimization flowchart**

The VPCRO method was proposed in [25]. It is based on the algorithm which was firstly introduced by Dijkstra [39] and was later utilized for different applications including optimal routing in transportation networks. The theoretical details of this method are throughly introduced in [25]. The transportation network is represented by a graph, consisting of $n$ nodes, $n_i \in \mathbf{N}$, where $\mathbf{N}$ is the set of all nodes in the network.

Transportation network graph is directed and the segment, $s_{ij} \in \mathbf{S}$, connects node $n_i$ to $n_j$, where $\mathbf{S}$ is the set of all segments in the network. We use Shanghai transportation network for the case study where the set $\mathbf{N}$ consists of 352 nodes denoted by $n_i$ and each node is assigned an index number *i* and associated with the corresponding 2-D coordinates collected from Google Maps. The 2-D coordinates are shown as longitude $n_i^{\text{lon}}$ and latitude $n_i^{\text{lat}}$ in unit meters for the node $n_i$. The set $\mathbf{S}$ contains 615 segments $s_{ij}$ of the network. Each segment is associated with weight information $w_{s_{ij}}^d$ and $w_{s_{ij}}^c$ which correspond to the distance, $d_{s_{ij}}$, and the transportation cost of the segment in terms of energy consumption, respectively.

We choose the traffic conditions on Monday 8:30 a.m. for city of Shanghai which is categorized into three states: heavy, normal and low traffic jam. These traffic conditions can be mimicked by certain driving patterns since traffic flow organizes the speed of the vehicle over time. It is assumed that the traffic conditions of each segment can be approximated by certain driving patterns. For low traffic conditions, Highway Fuel Economy Test (HWFET) driving cycle is employed which means that the road condition is similar to driving on highway. For normal traffic conditions, Urban Dynamometer Driving Schedule (UDDS) driving cycle is chosen. For the segments with high traffic, the New York City (NYC) driving cycle is used since it reflects the conditions of driving in heavy traffic just like New York City with frequent stop and go.

For our study, we have considered four types of electrified vehicles: PHEVs with three different battery sizes and a Battery Electric Vehicle (BEV) [26]. PHEV*x* stands for a PHEV that can cover *x* miles by using only electricity. After *x* miles, PHEV switches to charge sustaining mode which means that gasoline is used as the energy source to cover the rest of the trip. BEV100 stands for a BEV that can travel 100 miles on electricity before needing a recharge.

Range extended PHEVs can be operated in two modes: the charge-depleting mode (CD) and charge-sustaining mode (CS). CD refers to a mode of electrified vehicle operation which relies on energy from the battery. Once the battery is depleted reaching at the target state of charge, the electrified vehicle changes its operation mode to CS, in which gasoline is used to provide all net propulsion energy and the electrical energy is utilized only as momentary storage to improve fuel economy [26]. The vehicle efficiencies under different driving patterns comes from the literature for CD and CS operation modes and details of simulations can be found in [26].

**Table 1: The specifications of the electrified vehicles used for the case study**

| Vehicle Type | Symbol | Unit | HWFET | UDDS | NYC |
|---|---|---|---|---|---|
| PHEV20 | $\mu_{CD}$* | mi/kWh | 5.7 | 6.2 | 4.2 |
|  | $\mu_{CS}$ | mi/gal | 58.6 | 69.4 | 45.7 |
| PHEV40 | $\mu_{CD}$ | mi/kWh | 5.7 | 6.0 | 4.1 |
|  | $\mu_{CS}$ | mi/gal | 58.2 | 68.0 | 43.1 |
| PHEV60 | $\mu_{CD}$ | mi/kWh | 5.6 | 5.7 | 3.8 |
|  | $\mu_{CS}$ | mi/gal | 57.8 | 65.8 | 40.3 |
| BEV100 | $\mu_{CD}$ | mi/kWh | 4.8 | 5.2 | 3.1 |

*$\mu$ is the efficiency of the vehicle under specific driving conditions

The distance between two nodes $n_i$ and $n_j$ on the transportation network is defined by the equation below:

$$d_{s_{ij}} = \begin{cases} C_d \sqrt{(n_i^{\text{lon}} - n_j^{\text{lon}})^2 + (n_i^{\text{lat}} - n_j^{\text{lat}})^2}, & \text{If } n_i \text{ and } n_j \text{ are connected} \\ \infty, & \text{Else} \end{cases} \quad (2)$$

where $C_d = 62.137$ mile/m is the scaling factor used to convert latitude and longitude data gathered from Google Maps from meters to miles.

Cost of each segment $c_{s_{ij}}$ on the transportation network depends on the vehicle type (PHEVx or BEV), unit energy cost ($p_{gas}$ and $p_{ele}$ for gasoline and electricity, respectively), vehicle efficiencies ($\mu_{CD}$ and $\mu_{CS}$ under CD mode and CS mode, respectively) and the remaining available battery energy of electric vehicles $E_{s_{ij}}^{rem}$ when they enter the segment $s_{ij}$. We give the cost equations for an arbitrary segment $s_{ij}$ for different kinds of electrified vehicles below.

$$w_{s_{ij}}^c = \begin{cases} p_{gas} \dfrac{d_{s_{ij}}}{\mu_{CS}}, & \text{if } E_{s_{ij}}^{rem} \leq 0, \text{for PHEVs} \\ p_{ele} \dfrac{d_{s_{ij}}}{\mu_{CD}}, & \text{if } E_{s_{ij}}^{rem} \geq \dfrac{d_{s_{ij}}}{\mu_{CD}}, \text{for PHEVs} \\ \left( p_{ele} E_{s_{ij}}^{rem} + p_{gas} \dfrac{d_{s_{ij}} - \mu_{CD} E_{s_{ij}}^{rem}}{\mu_{CS}} \right), & \text{else for PHEVs} \\ p_{ele} \dfrac{d_{s_{ij}}}{\mu_{CD}}, & \text{for BEVs} \end{cases} \quad (3)$$

$E_{s_{ij}}^{ini}$ is the initial available energy which is calculated by using the sensory information from the battery (battery capacity, current SOC and battery swing) at the beginning of the segment $s_{ij}$ [25][26]. The remaning energy in the battery after leaving segment $s_{ij}$ is denoted by $E_{s_{ij}}^{rem}$. For the road segment between node $i$ and $j$, the relation between these values can be represented as (4).

$$E_{s_{ij}}^{rem} = E_{s_{ij}}^{ini} - \frac{d_{s_{ij}}}{\mu_{CD}} \quad (4)$$

The calculation of least cost path is much more complicated due to the trade-off between using motor or engine, which use gasoline and electricity that has different costs, and also factors like traffic conditions, vehicle powertrain type and initial battery SOC. The main difference for energy consumption between these vehicles is the efficiency of vehicles when they are driving under different road conditions. We can define the path-finding problem as an optimization problem with the following objective functions.

$$\text{Least travel cost: } min \left( \sum_{(i,j) \in R} w_{s_{ij}}^c \right) \quad (5)$$

where **R** is the set of possible paths from the initial node to the goal node, $i$ and $j$ are the indices of the nodes in the path. Using the algorithm and methods in [25], we determine the least cost paths for each type of vehicles for each Origin-Destination (O-D) pairs. In addition to the objective function in (5) which is used for least cost optimal routing, we also consider the expected charging cost of each EV at each candidate charging station.

### 2.3.1. Breadth first search algorithm

When the travel path is determined for drivers using VPCRO [25], Breadth-first-search (BFS) algorithm [41] is applied to search nearby charging stations for drivers. BFS is a method for exploring a graph/tree data structures. It begins at the tree root (initial point for the EV drivers) and explores the neighbor nodes first, before moving to the next level neighbors. For each node among the travel path, we find the nearest charging station to reach one located on the optimal route. Then we choose five candidate charging stations among these cases. Offering several charging stations helps the drivers to have more options to decide for charging their battery. After finding five nearest charging stations, we proceed with choosing the charging station with least cost.

For the searching part, the beginning point is the passing node along the whole path of the trip. We use (6) which applies the equations of distance, the scaling factor of the search radius (k), and the search radius denoted by $d_{sea} = 100\ m$. $n_i^{lon}$ and $n_i^{lat}$ are the longitude and latitude of the current node on the obtained optimal route, and $k = 1$. Then we can get the longitude and latitude range of charging stations for the first searching area. If there is no qualified charging station then $k = k + 1$ until we find out the first charging station candidate for this passing node. The level of cost effectiveness of the set that contains the possible charging stations for each node depends on the search radius and the number of charging stations considered as candidates. Equation (6) verifies that whether the corresponding charging station is located in the search area. $n_{ch_j}^{lat}$ and $n_{ch_j}^{lon}$ denote the latitude and longitude of the charging station $j$. Furthermore, $C_d = 62.137$ mile/m is the scaling factor used to convert latitude and longitude data gathered from Google Maps from meters to miles.

$$C_d \sqrt{(n_i^{lon} - n_{ch_j}^{lon})^2 + (n_i^{lat} - n_{ch_j}^{lat})^2} \leq k \times d_{search} \tag{6}$$

### 2.4. Solving Optimal Power Flow Problem for Locational Marginal Electricity Prices

Balancing between load and supply is the main task of the independent system operator (ISO) to find the generation levels given the predicted load [42][43]. OPF problem is solved to determine the economically-optimal operating points of the power systems and to calculate the output power of each generator [42][44]. In order to use DC optimal power flow (DCOPF) we need to make some assumptions [45][46] such as: voltage angle differences being small, line resistance being negligible, and voltage profiles being flat. After solving DCOPF problem, ISO determines the locational marginal prices (LMPs). LMPs can be used as a pricing index to determine the value of electricity at each node [47][48]. It has been widely-used for energy pricing to reflect the value of electricity at each node.

In order to minimize the total power system generation cost, we solve DCOPF problem [43]. The updated load demand (compared with conventional load demand without EV charging demand) includes the EV charging demand. Hence, in order to achieve an accurate EV charging demand estimation, we consider EVs' routing optimization problem. To this end, we add the total hourly-estimated charging demand of EVs to the demand. Then, we solve the DCOPF problem, which is a minimization problem of the general form shown in (7).

$$\min_{X} f(X) \tag{7}$$
$$s.t. \ g(X) = 0$$
$$h(X) \leq 0$$

In this problem, the control variables are the active power outputs, while the state variables are the voltage angles at the buses. Given the following notations for the parameters and variables, we define the optimization problem as (8).

$\Omega_{sys}$: Set of buses in the system; $\Omega_G$: Set of all generators; $\Omega_{G_i}$: Set of generators at the $i^{th}$ bus; $\Omega_{L_i}$: Set of lines in the system incident on $i^{th}$ bus (all buses connected to bus $i$); $\Omega_D$: Set of buses to which load is connected; $P_{G_i}$: Output of generator $i$; $P_{D_i}$: Demand at bus $i$; $C_i(P_{G_i})$: Quadratic cost function of generator $i$; $B_{ij}$: The element in the susceptance matrix in $i^{th}$ row and $j^{th}$ column; $\theta_i$: Voltage angle at the $i^{th}$ bus; and $F_{ij}$: Flow in the line joining $i^{th}$ bus and $j^{th}$ bus.

$$\min_{P_{g_i}, \theta_i} \sum_{i \in \Omega_G} (a_i + b_i P_{g_i} + c_i P_{g_i}^2) \tag{8.1}$$

$$s.t. \sum_{k \in \Omega_{G_i}} P_{g_i} - P_{L_i} - \sum_{j \in \Omega_{L_i}} B_{ij} \times (\theta_i - \theta_j) = 0, \quad \forall i \in \Omega_{sys} \tag{8.2}$$

$$P_{G_i}^{min} \leq P_{G_i} \leq P_{G_i}^{max}, \quad \forall i \in \Omega_G \tag{8.3}$$

$$|B_{ij} \times (\theta_i - \theta_j)| \leq F_{ij}^{max}, \quad \forall i, j \in \Omega_{sys} \tag{8.4}$$

$$\theta_1 = 0 \tag{8.5}$$

where (8.1) is the total cost of generation of the entire system, (8.2) shows the nodal power balance of all buses, (8.3) enforces the generation limits, (8.4) represents the line flow limits, and (8.5) ensures the fact that slack bus has zero voltage angle. The Lagrange function for the general form shown in (7) is defined as (9).

$$\mathcal{L}(X, \lambda, \mu) = f(X) + \lambda^T g(X) + \mu^T h(X) \tag{9}$$

where $\lambda$ and $\mu$ are the Lagrange multiplier vectors for equality and inequality constraints respectively and $g(X)$ and $h(X)$ are set of equality and inequality constraints, respectively.

The Lagrange multiplier vector, which corresponds to the equality constraint, includes the negative value of the LMP at each bus. The LMP values are then used as the inputs for the charging station pricing strategy. We assume that all charging stations at each region are connected to an electrical power bus. This corresponding bus broadcast the obtained electricity price to its connected charging stations. The loop is closed here, i.e. charging stations communicate with EVs regarding the electricity bids and update the expected demand of each EV. After performing one iteration, the expected electricity demand is broadcasted to power system operator which is used to solve the optimal power flow problem and update the LMPs iteratively.

## 3. Case Study and Simulation Results

In this section, we present our results. To this end, first we introduce the analyzed power and transportation network topologies. Then, we evaluate the power systems and transportation network operation with and without using our proposed framework and methodologies.

### 3.1. Proposed Power and Transportation Network Test Systems

The one line diagram of modified IEEE 9-bus test system before load demand modifications is depicted in Fig. 8 and updated parameters are shown in Table 2. As shown in the figure, there are six regions for charging stations. These regions are connected to the power system using six distinct buses.

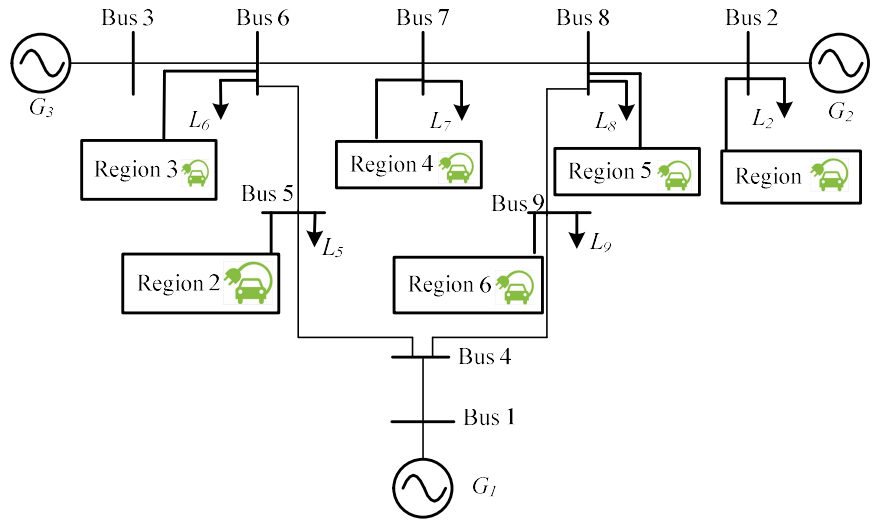

**Fig. 8. One line diagram of modified IEEE 9-bus test system**

We modify the load demand values at the load points. The updated values are shown in Table 2. This table also includes the maximum number of EVs that can be served (parked and charged) in each region. Charging stations at each region are connected to the power network via specific load points as shown in Fig. 8.

**Table 2: Load demand and maximum capacity of EV charging stations at each bus**

| Bus number | 1 | 2 | 3 | 4 | 5 | 6 | 7 | 8 | 9 |
|---|---|---|---|---|---|---|---|---|---|
| Load demand (MWh) | 0 | 200 | 0 | 0 | 120 | 10 | 160 | 40 | 80 |
| Region index | NCS* | 1 | NCS | NCS | 2 | 3 | 4 | 5 | 6 |
| Max charging station capacity (# of EVs) | NCS | 500 | NCS | NCS | 500 | 700 | 600 | 500 | 1500 |
| Number of charging stations | NCS | 7 | NCS | NCS | 5 | 8 | 6 | 8 | 16 |

*NCS: No charging station

The number of EVs in China is estimated to be 0.3 million in 2015 [49]. 15% of these vehicles are estimated to be used in Shanghai [49]. In order to obtain a practical test system, we evaluate our proposed framework by considering 45000 EVs. The details of the modeling of the Shanghai transportation network is given in Section 2.3. Fig. 9 illustrates the constructed traffic network for the simulation purposes based on Google map information of Shanghai, China, and Fig. 10 shows the charging stations in each region based on their electrical connection point to the power network. The light, medium and heavy traffic conditions are denoted with three colors: green, orange, and purple, respectively and charging stations are represented with squares.

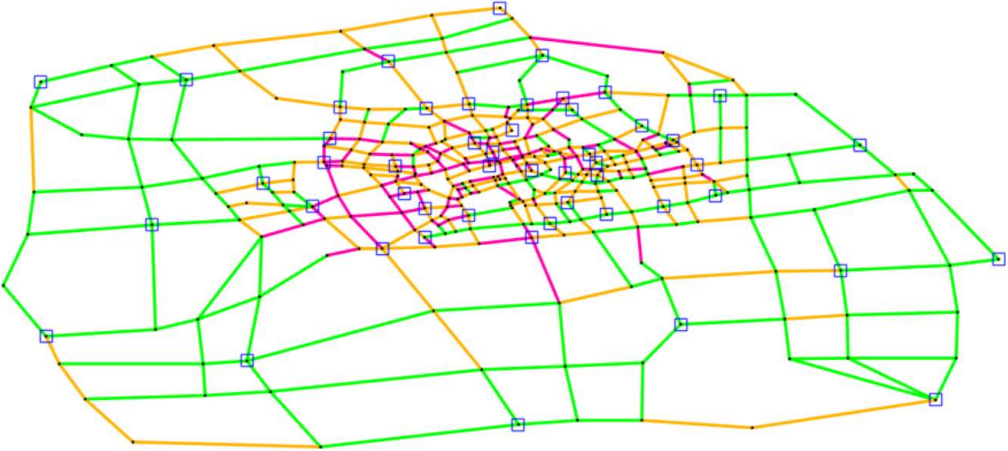

**Fig. 9. Topology of analyzed Shanghai transportation network with charging stations**

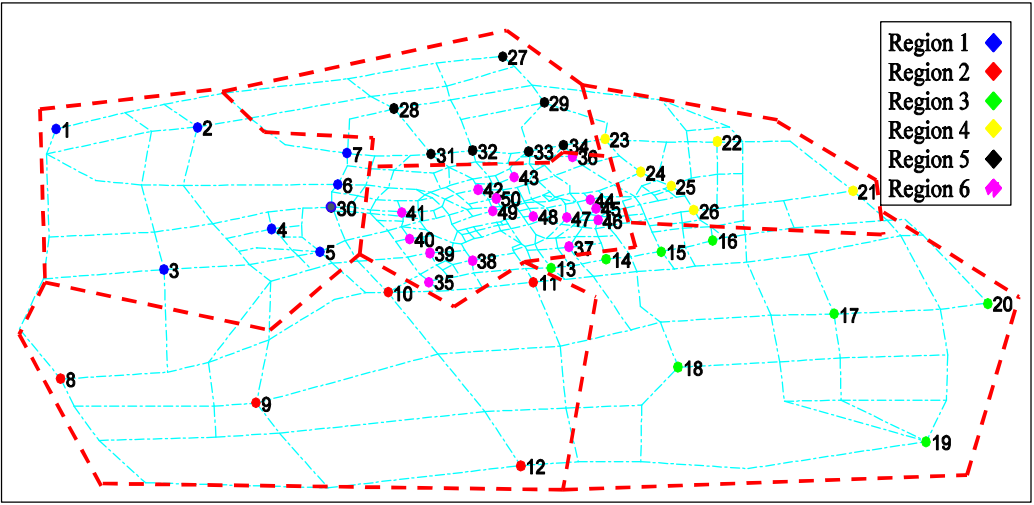

**Fig. 10. Distribution of the charging stations in specified geographical regions**

We use randomly-generated stochastic Origin and Destination (O-D) pairs to simulate traffic situations. Furthermore, we consider the penetration level of PHEV20, PHEV40, PHEV60, and BEV100 to be 16%, 16%, 16%, and 52%, respectively. The distribution of utilized *CPI* values for all of the 50 charging stations is presented in Fig. 11. In this study we assumed that the efficiency of charging stations is 100%.

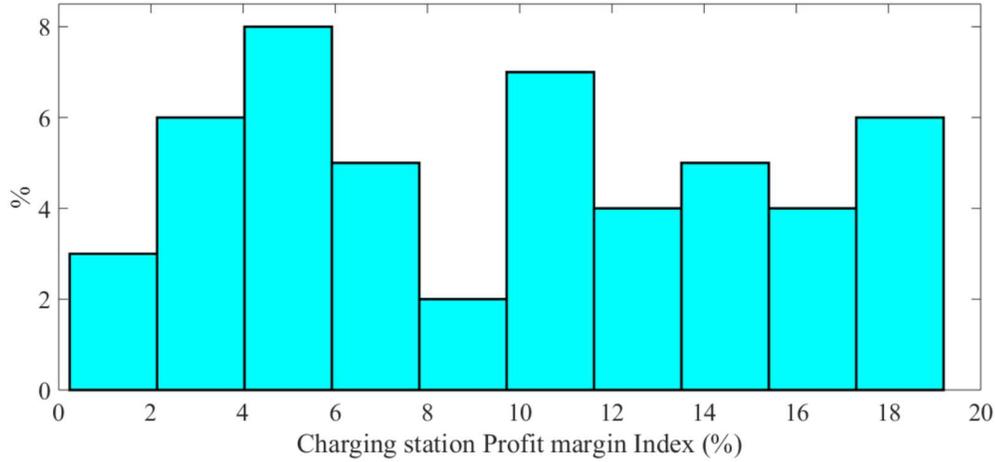

**Fig. 11. Distribution of the profit margins for the charging stations**

## 3.2. Power Systems and Transportation Network Operation With and Without Using the Proposed Framework

In order to illustrate the effectiveness of our proposed framework, we compare the optimal routes for three randomly chosen trips in our case study. Further, we find the total EV charging cost of each region for both scenarios: with and without the proposed methods. Finally, the electricity price values and total power system operation cost is compared for all of the vehicles.

### 3.2.1. Impact of the simultaneous optimization on vehicle routing

Fig. 12 shows the impact of our proposed methodology for the optimal joint operation of transportation and power networks on the cost-optimal routes of randomly selected three electrified vehicles from our case study. Columns of the Fig. 12 show (1) the traffic conditions of the transportation network, (2) cost-optimal routing result using VPCRO and finally (3) cost-optimal routing result using CSS-VPCRO from left to right. Rows 12.a, 12.b, and 12.c show the routing results for three different EVs. The blue points represent the available charging stations and the squared blue points represent the selected charging station on the cost optimal path. For the first randomly selected EV (Fig. 12.a) in the first scenario that is shown in the middle map, charging stations 39, 40, and 41 have the same electricity price (5 cents/kWh) because they are located in the same region. Thus, *CPI*=0% for all charging stations. VPCRO chooses charging station 41 since it is in the proximity of the cost optimal route and closer to destination. The reason for not choosing charging stations 39 and 40 (while having the same charging cost as 41) is trying to obtain the maximum achievable SOC at the destination. If the EV is charged at charging station at 39 or 40 to a certain level, the battery SOC will be less compared to the case of charging at 41 to the same SOC. Given similar options with the same price, it is prefered to choose the one that provides us with the maximum SOC at the destination. In the second scenario for the first EV (right map), the CPI values for charging stations 6, 39, 40, 41, and 30 are 9.9%, 10.8%, 11.3%, 7.6%, and 1.1%, respectively. The new routing result shows that although there are three charging station located on the optimal route, our algorithm selected charging station 30 as the optimal one to purchase electricity. Charging station 30 used an incentive to attract the EV to purchase electricity from them. This changed the optimal route. Similar explanations hold for the second EV (Fig. 12.b) and it is shown that our strategy CSS-VPCRO might change the optimal routes of EVs. For the third EV, the *CPI* values of charging stations 33 and 24 in the second scenario are 1.5% and 9.2%, respectively. Hence, the routing algorithm chooses charging station 33 rather than the one on the least cost path. Results show that interaction of EVs with charging stations during vehicle routing changes the optimal routes.

Fig. 12. Simulated paths with different routing strategies and electricity price signals:

**Vehicle Power Train Connected Routing Optimization (VPCRO) and Charging Station Strategy-VPCRO (CSS-VPCRO)**

**3.2.1. Impact of the simultaneous optimization on EV charging demand and cost**

Figures 13 and 14 show the total charging demand and cost for 6 regions under two scenarios: 1) without charging station pricing strategy under fixed electricity price signals from power systems, i.e. the results of DCOPF does not affect the optimal route, and 2) with charging station pricing strategy and considering electricity price signals based on DCOPF.

The proposed optimal operation strategy of the interdependent power and transportation networks reduces the total charging demand, as well as the total transportation cost of each region, as shown in Figures 13 and 14. It also validates the performance of our strategy in terms of motivating EVs to use the cost-effective routes which passes through the charging stations with lower price signals.

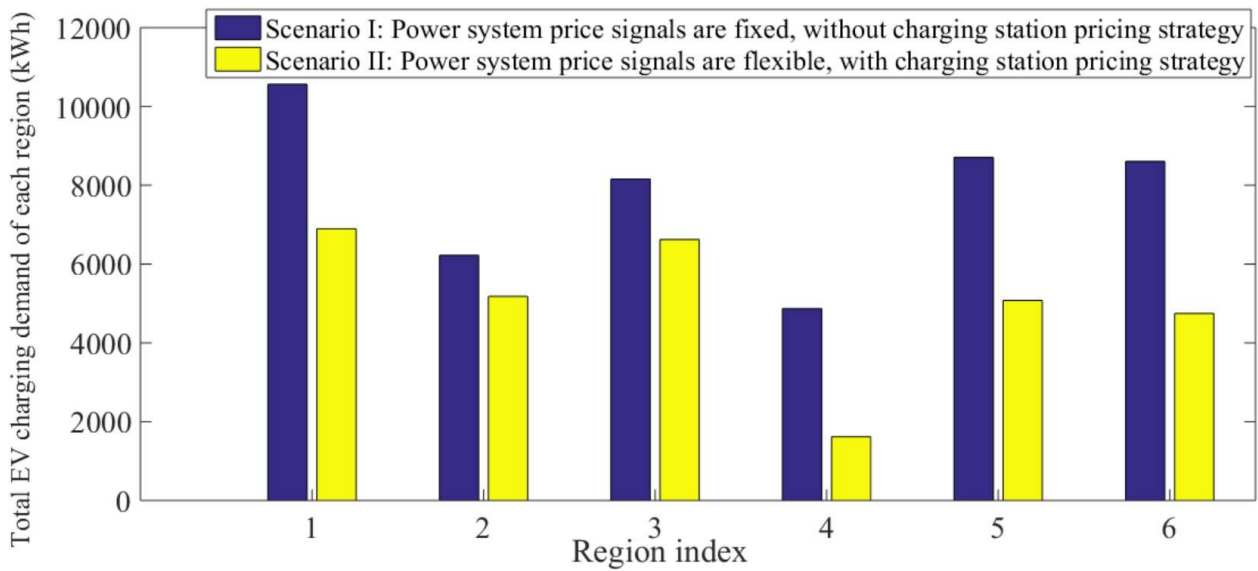

**Fig. 13. Total hourly charging demand of each region under various scenarios**

Fig. 13 shows that the fixed price signals without charging station pricing strategy (Scenario I) will lead to higher charging demand compared to the case where the charging stations modify the price signals to attract the EV drivers. Furthermore, for Scenario II (with charging station pricing strategy), after considering the price variations of power systems, the charging demand of all regions will reduce. This reveals the effect of power system price signals on EV drivers to change their route to use more cost effective charging stations, as well as reducing their charging demand. For instance, power system price variation will reduce the electricity price in region 3. Consequently, more EVs will charge their battery in this region and it will increase the total charging demand of this region compared to others. Ignoring the effect of EV charging demand on the electricity price will deteriorate the optimal operation point of both power and transportation networks. Scenario I highlights the effect of fixed electricity prices (ignoring power systems reaction to load increase incurred by EV charging stations) on the total EV charging demand. Note that higher EV charging demand leads to the following consequences: 1) increasing the charging cost, and 2) increasing the total power system operation cost. One of the main advantages of our proposed framework is modeling the interdependency of power and transportation networks to obtain a more realistic optimal operating point for both networks: (1) Our framework considers the effect of EV charging demand on the electricity price in power systems, (2) It also models the effect of the electricity price variations on the optimal routing of EVs in transportation network by deploying the novel CSS-VPCRO routing approach.

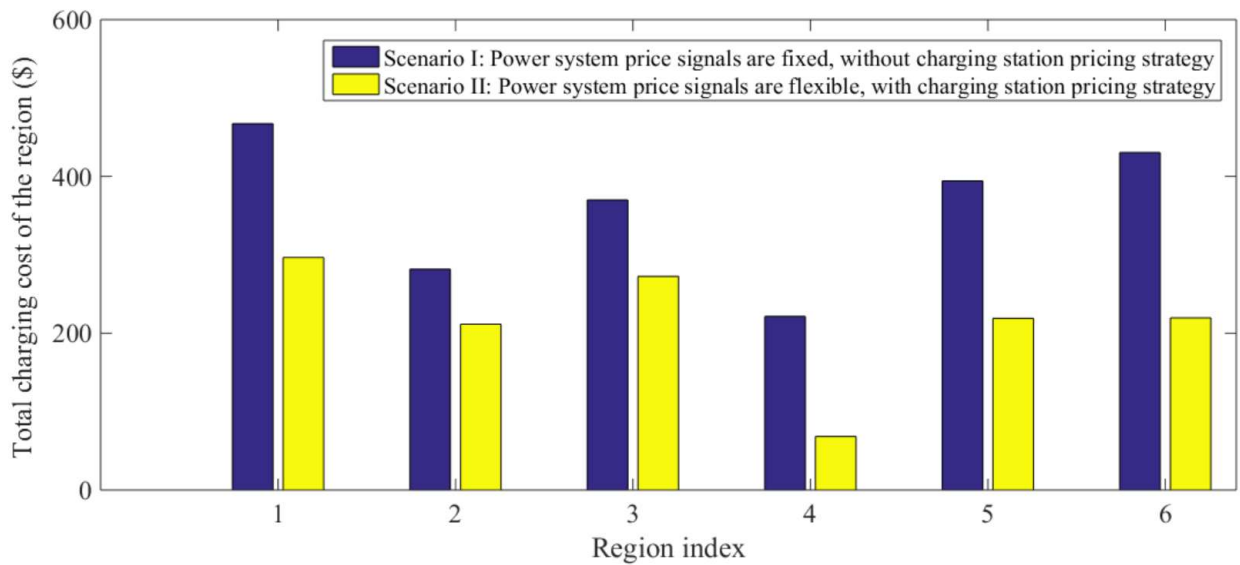

**Fig. 14. Total hourly charging cost for each region under various scenarios**

According to Fig. 14, fixed price signals without charging station pricing strategy will lead to higher charging cost compared to the case where the charging stations change the price signals and power system updates the electricity price at each region according to the EV charging demand. In other words, if we ignore the dynamic nature of power systems which leads to price variations based on EV charging demand, we will have different optimal operating point for both power systems and transportation networks. Furthermore, after considering the price variations of power systems, the charging cost of all regions reduce for Scenario II (with charging station pricing strategy). Different regions have different amount of cost reduction based on their electricity price. This is due to the fact that each charging station has a different pricing scheme to attract EV drivers. In other words, when there is a cheaper charging station in another region, EVs might prefer to choose a more energy-consuming route to charge their battery more economically. EV drivers might choose a longer distance which leads to a lower charging cost due to using a cheaper charging station. Power systems reaction to the load change due to EV charging demand can significantly affect electricity prices, and eventually the routing of EVs when we consider the communication of EVs with competing charging stations.

Our analysis validates the effectiveness of the proposed framework in terms of reducing total power systems cost as well as transportation network related charging costs as shown in Table 3. Total power system cost refers to the cost of generation to supply the total power system load demand. It also includes the additional cost incurred by integrating electric vehicle charging stations into the power networks. Note that we consider the total payment of all EV drivers corresponding to electricity charging cost as the total transportation cost. The total power systems cost due to conventional load before integration of EV charging is $15,308 per hour (base case) in our case study.

**Table 3: Comparison between total power system cost and transportation network cost for different scenarios**

| Scenario | I | II |
|---|---|---|
| Total power system cost ($/hr.) | 17,476 | 16,677 |
| Total charging cost ($/hr) | 2,165 | 1,287 |
| Additional power system cost (compared to the base case) | 14.1% | 8.8% |

As shown in Table 3, our proposed framework for optimal operation of power and transportation networks reduced the additional power system cost introduced by the EV charging demand up to 38% compared to base case. It reduced total power system cost by 5% and total charging costs by %40. In other words, using a EV-charging station communication and bidding mechanism while finding the optimal routes will reduce both power system and transportation network costs occurred by electricity consumption and generation.

## 4. Conclusions

In this paper we proposed a novel framework for interdependent power and electrified transportation networks. To this end we proposed an iterative least cost vehicle routing process that utilized the communication of electric vehicles (EVs) with competing charging stations to exchange information, such as electricity price, energy demand, and time of arrival. We called this routing strategy as "Charging Station

Strategy - Vehicle Powertrain Connected Routing Optimization (CSS-VPCRO)". In order to find the electricity price at each region we solved the DC optimal power flow problem. Charging stations communicate the updated electricity prices (based on their pricing strategy and locational marginal prices from optimal power flow) to EVs. Nodal electricity price acts as the shared parameter among the two optimization problems, i.e. optimal power flow and optimal routing problem. Electricity price depends on power demand which is affected by charging of EVs. Location of EV charging stations and their different price schemes might affect the routing decisions of EVs.

The effectiveness of the proposed approach is evaluated using Shanghai transportation network and IEEE 9-bus test system. We divided the transportation network into 6 regions. Each of these regions assumed to be supplied from one of the load points of the IEEE 9-bus test system. The simulations results for 45,000 EVs validates the outperformance of the proposed framework in terms of both total cost and total energy demand. The additional cost incurred by electrifying the transportation network is reduced from 14.1% (for the conventional routing schemes) to 8.8% for the proposed simultaneous optimization of power and transportation networks. Consequently, the simulaion results validated the performance of our framework in two main directions: (1) reducing the electrified transportation cost for the analyzed electrified transportation network by 40%, and (2) reducing the total power system cost by 5%. Our novel approach that combines the electrified transportation with power system operation holds tremendous potential for solving electrified transportation issues and reducing energy costs.

## Acknowledgements

The authors would like to acknowledge the valuable inputs and fruitful comments from Prof. Marija D. Ilić from Carnegie Mellon University to this article. The main idea of this paper has been substantially improved based on her advantageous inputs. We would also like to thank Zhiqian Qiao from SYSU-CMU Joint Institute of Engineering, for her help with the transportation system simulation and members of the Laboratory for Intelligent Vehicles and Energy Systems (LIVES) for their comments.